\documentclass{article} 

\usepackage{graphicx}

\title{Tipping point analysis of  electrical resistance data 
with early warning signals of failure for predictive maintenance}

\author{V.~N.~Livina$^1$\footnote{Corresponding author, email:valerie.livina@npl.co.uk}, A.~P.~Lewis$^1$ and 
M.~Wickham$^1$ \\
{\it $^1$National Physical Laboratory, Teddington, 
United Kingdom}}

\date{}

\begin{document}

\maketitle

\begin{abstract}
We apply tipping point analysis to measurements of electronic 
components commonly used in applications in the automotive or 
aviation industries and demonstrate early warning signals based 
on scaling properties of resistance time series. The analysis 
is based on  a statistical physics framework with a stochastic 
model representing the system time series as a composition 
of deterministic and stochastic components estimated 
from measurements. The early warning signals are observed much 
earlier than those estimated from conventional techniques, 
such as threshold-based failure detection, or bulk estimates 
used in Weibull failure analysis.
The introduced techniques 
may be useful for real-time predictive 
maintenance of power electronics, 
with industrial applications. We suggest that this approach 
can be applied to various electric measurements in power 
systems and energy applications.
\end{abstract} 

\section{Introduction}

A significant challenge in the design and development of 
high-reliability electronic assemblies is in relating the 
results from testing  to real-world performance. The key 
issue is that, due to the wide range of applications, 
standardised tests may not predict the range of harsh 
environments to which an electronic assembly will be exposed. 
As a consequence for high-reliability use, particularly in 
safety-critical applications, a common approach is to use a 
standardised test which utilises over stress conditions to 
accelerate failures. 

Weibull reliability analysis \cite{scholz} 
uses a parametric Weibull model to estimate 
a probability density and failure rate function based on 
the parameters of the distribution of parts failure, which 
provide information about the average behaviour of parts 
of expected quality. Factors such as time-to-first failure 
and analysis of Weibull distributions are used to give 
indications of when components/assemblies should be replaced. 

However, since there are significant uncertainties in real 
stress conditions and quality of materials and manufacturing process, 
these indicators are  based on the assumptions that 
are highly conservative:
 the Weibull plot gives the time to first failure 
(these are not conservative), however it is further used for 
stringent assumptions.

Therefore alternative methods are sought to give a more 
reliable indication of the remaining useful life. One method 
which has shown promise is the use of prognostic devices for 
monitoring solder joints \cite{chauhan}. 
These can be components (e.g., zero-ohm resistors) 
which are incorporated into an electronic assembly and 
are designed to fail before any other component. 
As they are a part of the assembly, they will experience 
the same manufacturing, environmental and detrimental factors. 
As part of a feasibility study on a range methods for measuring 
the progression of failure for a solder joint 
(between a printed circuit board --- PCB --- and a zero-ohm resistor), 
an experiment was conducted which looked at the evolution 
of the DC resistance as the solder joint underwent thermal cycling. 

The thermal cycling ageing process can cause cracks to form 
in the solder interconnect due to mismatches in the coefficients 
of thermal expansion between the substrate, contact pads, 
solder and components. These cracks cause discontinuities 
in the electrical circuit, although the resistance change 
during the crack initialisation tends to be minor until a 
failure event where the resistance increases significantly 
by several orders of magnitude, potentially to open circuit. 

The hypothesis we test is whether 
small changes in electrical resistance and in the pattern 
of their fluctuations 
(such as short- and long-term memory) can be used as early 
warning indicators to predict impending failure events. 
Early warning signals (EWS) based on these changes proved to be 
of general applicability in generic dynamical systems, and 
bringing them into the area of predictive maintenance may 
serve as a cross-disciplinary advantage for the manufacturing industry. 
Conventionally, early warning signals are statistical indicators
showing  the approach of critical 
transitions in a dynamical system, 
which is often hidden by system fluctuations. The logic behind 
applying these indicators for failure diagnostics is that EWS
indicators should be sensitive to changes 
in memory  (i.e., long-term dependencies quantified by 
auto-correlations) in the measurements
of devices, and this should happen earlier than 
any threshold-based transition is detected.

There  are a number of industrial 
guides that suggest various thresholds
and levels of tests (from stringent to moderate)
of electronic components with different properties. Some of them 
operate values of up to 10,000 ohms and are specific to particular 
devices. Triple-nominal resistance is an empirical threshold
that allows  us 
to identify  an abnormal increase of resistance before the 
hard open-circuit failure (which can be see as the vertical 
line in Figure~\ref{f2}). It is general enough to assess components with
various characteristics. The hard failure with open circuit 
occurs later than the triple-nominal threshold, and our techniques 
detect the drift of the resistance variable even earlier. 

This demonstrates that the proposed techniques are applicable to
various systems, and can further be tuned for earlier detection 
if more stringent tests are necessary. The particular value of the 
methodology is that it analyses individual measurments of 
 specific components and forewarns about individual failures rather than 
operates statistical averages.

Tipping points are critical transitions and bifurcations
in time series data that may lead to another existing state of the 
dynamical system (such as change from regular dynamics to a failure), 
or to appearance and disappearance of system states, 
which may be crucial to condition monitoring.  
Time series analysis techniques that allow one to detect 
such tipping points may provide tools for predictive 
maintenance of dynamical systems, which is of particular importance
in electronic devices that are related to control and safety.

Tipping points in dynamical systems have recently become a 
topic of high interest in the area of climate change; see, 
for example, \cite{lenton08}. 
Applications of the tipping point analysis have been found so far in 
geophysics 
\cite{livina07,livina10,livina11,livina12}
\linebreak 
\cite{prettyman18,prettyman19},
structure health monitoring 
\linebreak
\cite{livina14}, 
as well as in ecology (see \cite{dakos,scheffer} 
and references therein). There is a debate about various types of 
tipping \cite{ashwin} and false alarms~\cite{ditlevsen}, 
but for practical applications in industry, 
non-bifurcational transitions 
(without structural change of the dynamical system), 
may be as important as bifurcations and require adequate 
analytical tools and techniques of analysis. 
One of the advantages of the tipping point methodology 
is that it does not require extensive training datasets 
(unlike many other techniques of machine learning), 
and therefore can be useful in situations where there 
is limited operational data available, or where stress 
conditions are unknown.

A dynamical system with observed time series of measurements 
can be modelled by the following stochastic equation with state 
variable $z$ and time $t$:
\begin{equation}
\dot z =D(z,t)+S(z,t),                                      	
\end{equation}
where $D$ and $S$ are deterministic and stochastic components, 
respectively. The probability density of the system 
can then be approximated by a polynomial of even order 
(the so-called potential system, see~\cite{livina10}). 
The stochastic component, in the simplest case, may be 
Gaussian white noise, although in real systems it is often 
more complex, for example, with power-law correlations, 
multifractal and other nonlinear properties. 

Tipping points can be described in terms of the underlying system 
potential $U(z,t)$, whose  state derivative, if it exists, defines 
the deterministic term in Equation~(1), 
i.e. $D(z,t)=-U'(z,t)$ \cite{livina11}. If the potential structure (number of potential wells) changes, 
the tipping point is a genuine bifurcation. 
If the potential structure remains the same, while the trajectory 
of the system samples various states, such a tipping point 
is transitional. An example of such transition may be the 
record of global temperature, which has the same structure 
of fluctuations with a drift 
(under forcing or noise-induced). 
In practical terms, both transition and bifurcation may lead 
to catastrophic damage of devices, however, genuine 
bifurcations tend to have more gradual dynamics compared with 
abrupt transitions, and therefore are more likely to provide early 
warning signals.  
An example of a genuine bifurcation in time series is the appearing
or disappearing state of the system potential. This happens with 
gradual shallowing of a potential well, as shown in \cite{livina11}.

The methodology has general applicability for studying trajectories 
of dynamical systems of arbitrary origin and serves to anticipate, 
detect and forecast tipping points. In this paper we apply the first 
stage of the tipping point analysis, the early warning signals 
for anticipation of tipping points, which is based on degenerate 
fingerprinting~\cite{held} 
with further modifications of the technique using 
Detrended Fluctuation Analysis 
\cite{livina07} and power 
spectrum~\cite{prettyman18}.

\cite{livina07} and \cite{prettyman19} provided a number of 
simulation experiments to explain technical differences between 
these indicators. In practice, it is useful to apply several of 
them to evaluate their performance for a particular system, 
as it is done in this study.

The methodology of early warning signals is based on a generic 
stochastic model with a pseudo-potential\cite{held,livina10}, which 
describes the states of the dynamical system and their evolution. 
The slowing down effect is a manifestation of the shallowing of the 
state potential well, or appearing new one, or drifting of the system
potential (this is further illustrated in \cite{livina11}). This
stochastic model is generic and applicable for approximation of 
dynamics of many systems, whose time series (trajectories) can be
studied with this methodology.

\section{Data}

A PCB test vehicle was designed to enable the evaluation of test 
methods to measure the remaining useful life of solder joints. 
To accelerate the ageing process, the test boards were placed in 
a thermal cycling chamber cycling from -55$^o$C to 125$^o$C 
at a rate of 10$^o$C min$^{-1}$ and with 5-minute dwells 
at the temperature extremes.  We use 5-minute dwells 
during the data collection from thermal cycling to ensure 
the whole assembly reaches the set temperature. 
Five minutes is a standard value to use  
(see, for example, \cite{jedec}, which uses values of 
the range 1-15 minutes).

The experimental setup monitored three measurement channels 
(each channel captured data from a 4-point probe measurement setup). 
Limitations on experimental equipment availability meant 
that we were required to run three separate experimental 
runs to collect the full dataset.

When measuring very low values of resistance 
(e.g. that of a solder joint), using the 4-point probe 
method is preferable (to remove any contribution to the 
measured resistance from the measurement leads). 
This means we can be confident that measured changes 
in the resistance are due to the degradation of the solder joint.

Thermal cycling induces failures at interfaces due to a mismatch 
in the coefficient of thermal expansion (CTE) at those interfaces. 
Therefore, whilst failure of the resistor would affect the 
outcomes significantly, it is highly unlikely to occur before 
solder joint failure. Further to this, if a resistor failed 
during the test, it would be picked up after the test when 
the resistors are checked to confirm they are still low resistance.

\begin{figure}[ht!]
\centerline{\includegraphics[width=0.6\columnwidth]{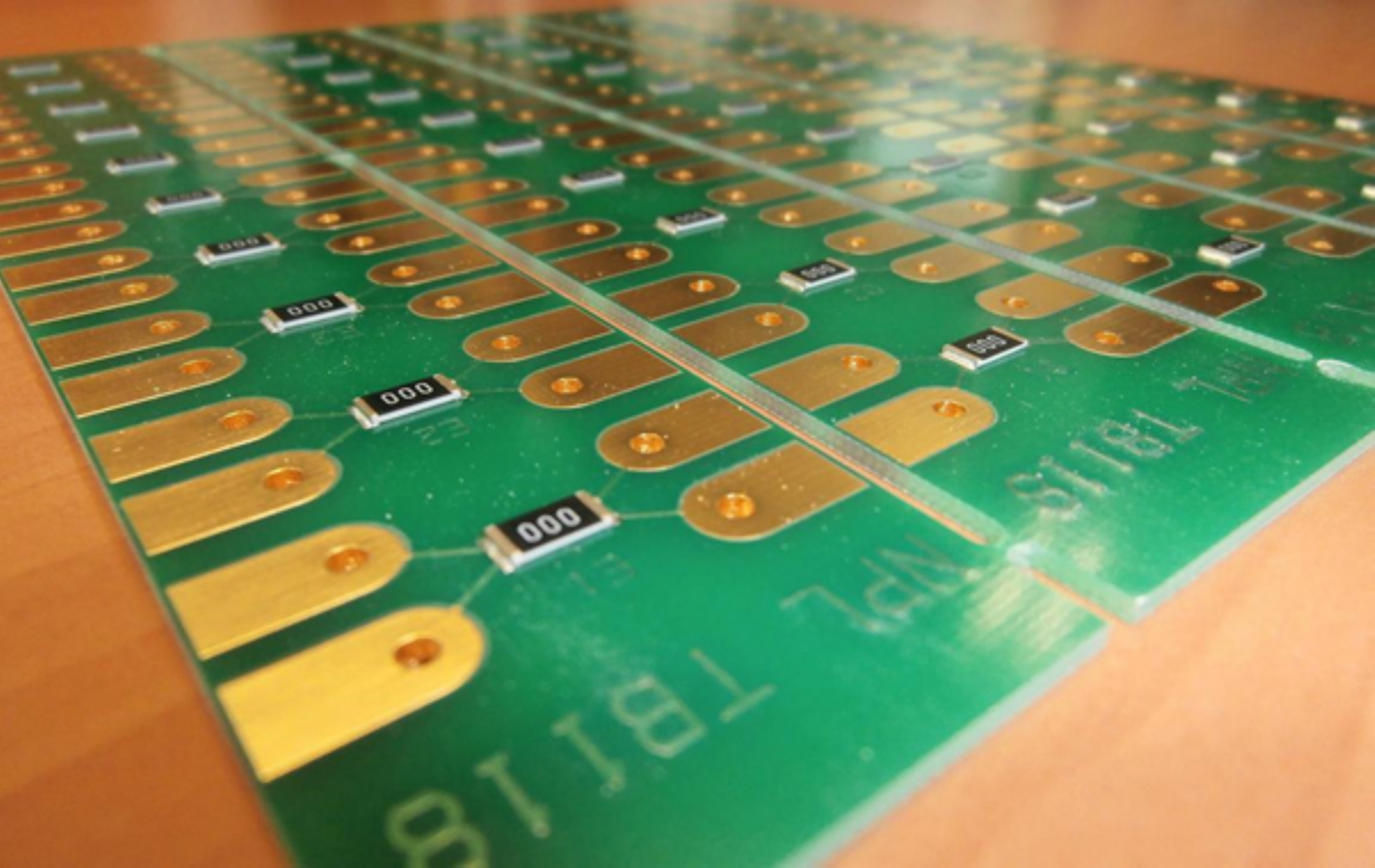}}
%\begin{center}
\caption{Image showing test boards. 
Each test board could host up to 10 surface 
mount chip resistors. 
Each component was connected to four pads to enable a 
4-point probe method of resistance monitoring.}
%\end{center}
\label{f1}
\end{figure}

\vspace{5mm}

The test boards were 1.6~mm 
thick copper clad FR-4 with a NiAu finish. 
The test components were zero-ohm 2512 chip resistors 
connected with a Pb-free solder interconnect. An image 
of the test board is given in 
Figure~\ref{f1}.

\begin{figure}[ht!]
\centerline{\includegraphics[width=1.\columnwidth]{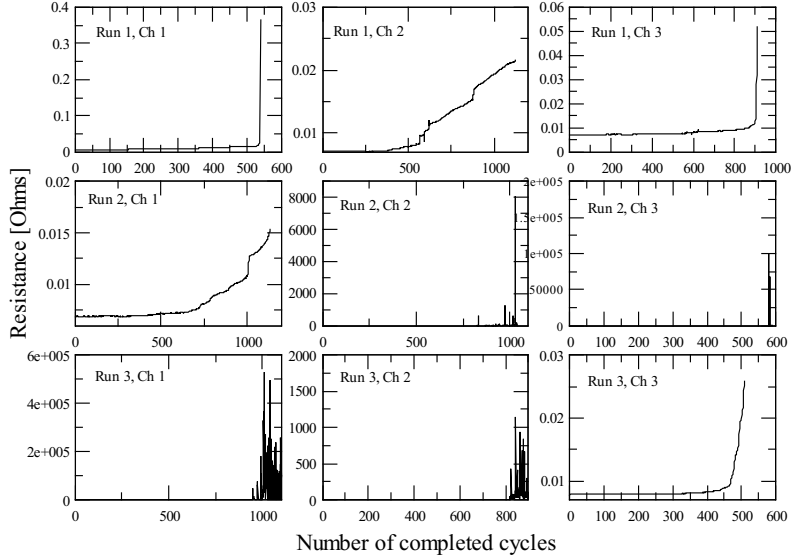}}
\caption{Nine time series of measured resistance with 
critical rise indicating the failure of the units after 
repeated test cycles 
(three runs using three channels denoted by 'Ch').}
\label{f2}
\end{figure}

We analysed measured resistance datasets from nine units, 
which experience failure 
(critical rise of resistance) after repeated testing cycles, 
see Figure~\ref{f2}.
 The reported cycles when the units went open circuit: r.1c.1 
(run 1 channel 1)
--- 540, r.1c.2 --- 1000, r.1c.3 --- 750, r.2c.1 --- 1000, 
r.2c.2 --- 815, r.2c.3 --- 810, r.3c.1 --- 910, r.3c.2 --- 543, 
r.3c.3 --- 516. 

\section{Methodology}

Anticipating tipping points 
(pre-tipping or early warning signal) is based on the 
effect of slowing down of the dynamics of the system. 
When a system state becomes unstable and starts a transition 
to some other state, the response to small perturbations 
becomes slower, 
which is often caused by the shallowing of 
the potential well \cite{livina11}.

This signal of ``critical slowing down'' 
is detectable as increasing autocorrelations quantified 
by the autocorrelation function 
(ACF) in the time series \cite{held}. 
Alternatively, the short-range Detrended Fluctuation 
Analysis (DFA)
\cite{livina07} or power spectrum (PS) scaling 
exponent 
\cite{prettyman18} can be monitored. 
These three techniques are essentially equivalent 
as they are monitoring the changes of ``memory''
(autocorrelations) in the data. 
 The main difference between ACF-indicator \cite{held} and 
DFA-indicator \cite{livina07} is that DFA has built-in detrending
procedure, which removes polynomial trends of various orders. 
We use DFA of order 2, which removes linear trend from the time 
series in sliding windows. This means that 
when comparing ACF- and DFA-indicators, we can attribute the 
differences between them to the presence of trends.
In the context of this work, for early warning signals both 
trends and increasing auto-correlations indicate destabilisation
of resistance time series and are important to detect for the 
purposes of predictive maintenance.
We explain in full 
the ACF-indicator technique, which is simplest of the three, 
and provide further references on DFA and PS-indicators for 
those who are interested in applying all three techniques for 
comparison. 
PS-indicator may require longer
datasets as power spectrum of short subsets may be affected
by noise~\cite{prettyman18,prettyman19}.

The early warning signal value is calculated in sliding windows of fixed length
(or variable length for uncertainty estimation) along a time series. These dynamically derived values form a curve of an early warning indicator whose pattern describes the behaviour of a time series. If the curve of the indicator remains flat and stationary, the time series does not experience any critical change 
(whether bifurcational or transitional). If the indicator rises to the critical value of one 
(the monotonic rise is assessed using Kendall rank correlation), it provides a warning of critical behaviour. 

Lag-1 autocorrelation is estimated by fitting an 
autoregressive model of order one (linear AR(1)-process) 
of the form\cite{held}:
$$
z_{t+1}=c\cdot z_t + \sigma \eta_t,
$$ 
where $\eta_t$ is a Gaussian white noise process of unit variance, and the ``ACF-indicator'' 
(AR(1) coefficient) is as follows:
$$
c= e^{-\kappa\Delta t}.
$$ 
where $\kappa$  is the decay rate of perturbations, $\Delta t$ is the 
time interval and $c\to 1$ as $\kappa\to 0$ while a 
tipping point is being approached. This analysis can 
be performed using several early warning indicators, 
for ACF -- with or without detrending data in sliding 
windows~\cite{livina12}.

\phantom{.}
\vspace{30mm}

\section{Results}

We have calculated two early warning indicators, ACF- and 
DFA-based, and compared the timing of the obtained early 
warning signals with the conventional threshold-based warning. 
As a threshold of failure, one can consider triple-nominal 
resistance. In the beginning of the experiment, the resistance 
values of the tested units were about 0.008 Ohm, and therefore 
the threshold would conventionally be established at about 0.025 Ohm. 

\begin{figure}
\centerline{\includegraphics[width=1.\columnwidth]{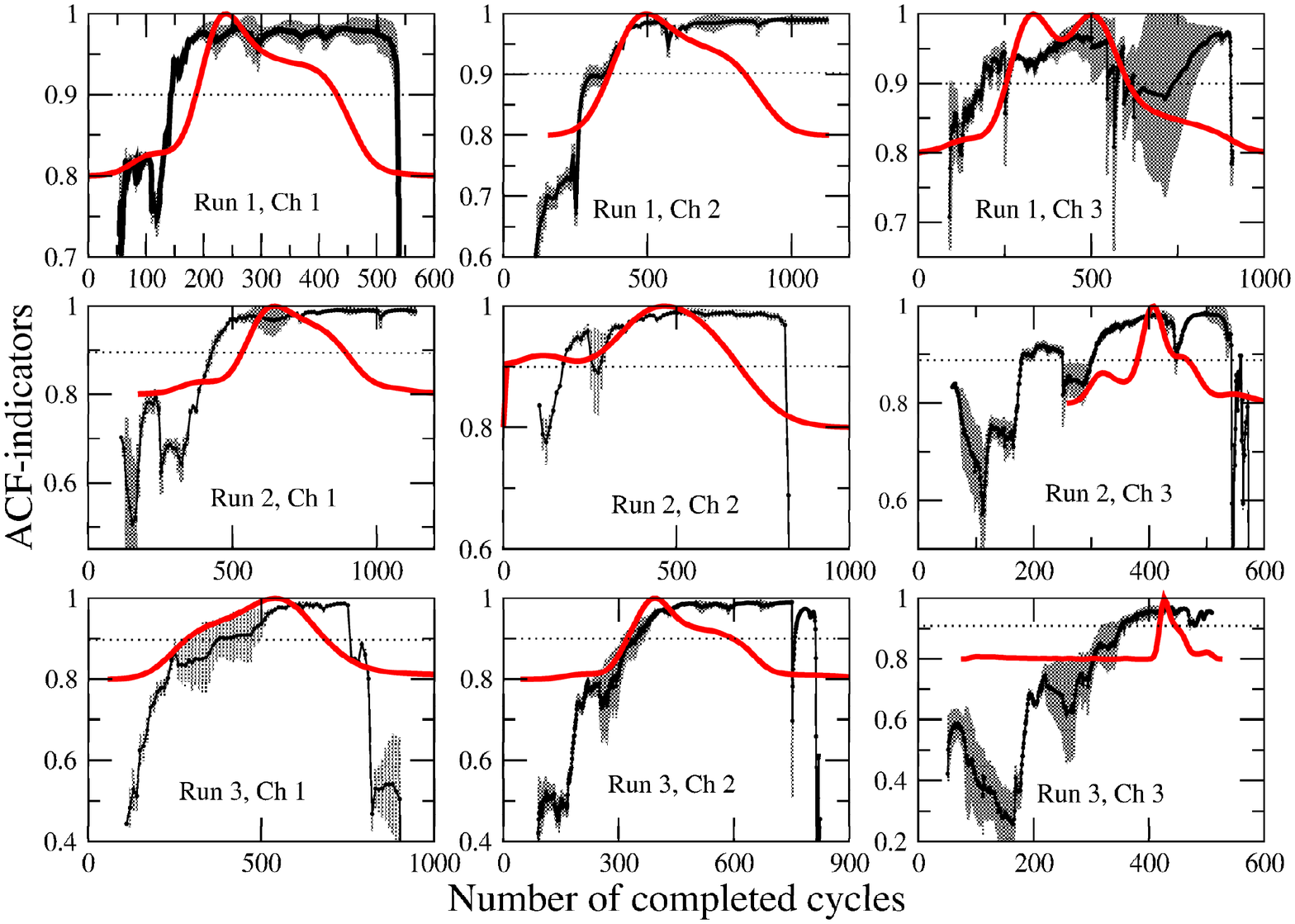}}
\caption{ACF-indicators with uncertainty quantification based on 
varying window size 
(10-50\% of time series length) for resistance data, 
with clear critical rise prior to failures. 
Dashed lines denote the high level of auto-correlation. 
Red curves correspond to probability 
densities (kernel distributions of the times of 
projections of subsets of the indicators when critical 
value 1 would be reached).
Panels correspond to the data in Fig.\ref{f2}.}
\label{f3}
\end{figure}

\begin{figure}
\centerline{\includegraphics[width=1.\columnwidth]{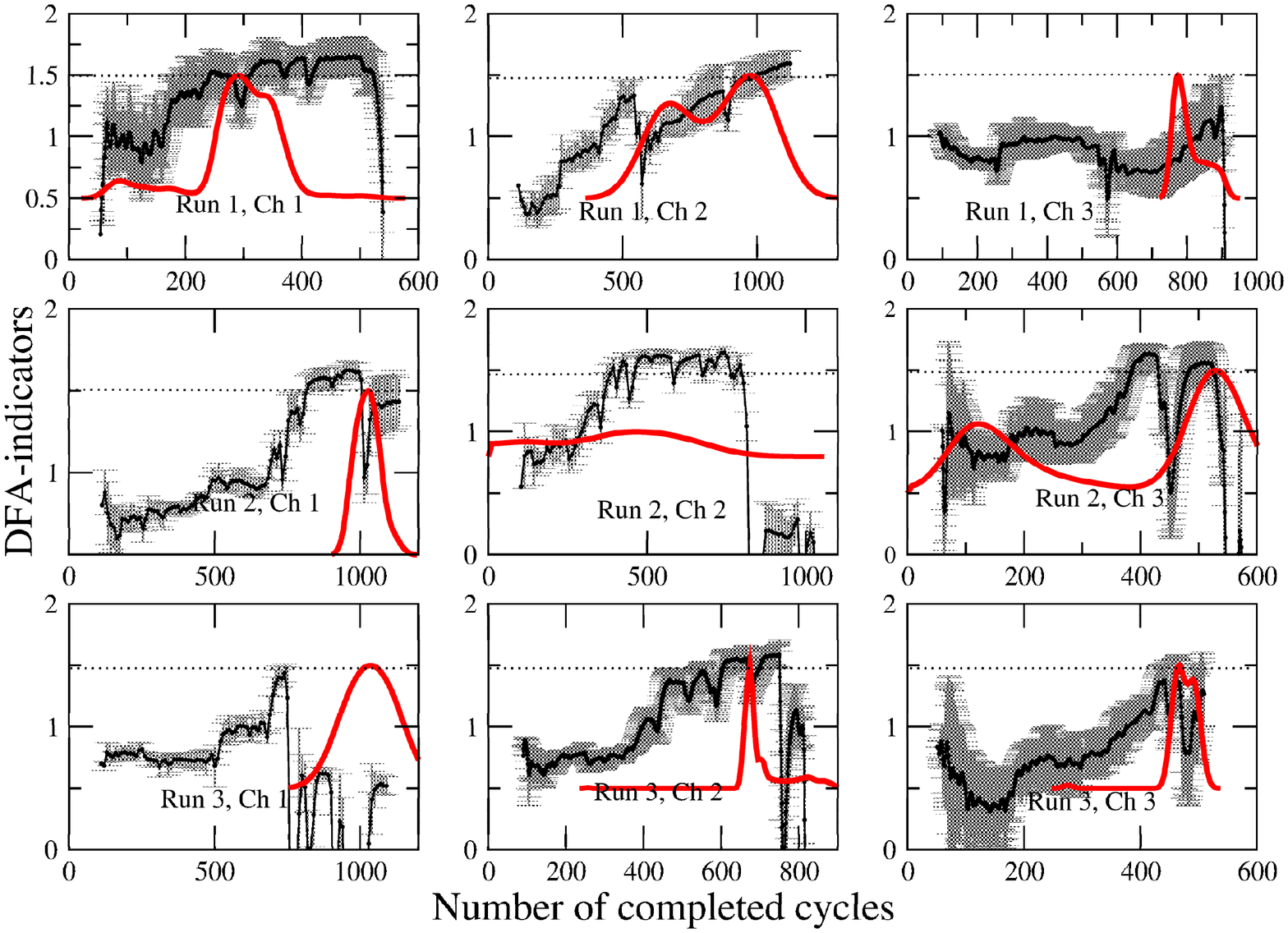}}
\caption{DFA-indicators with uncertainty quantification 
based on varying window size 
(10-50\% of time series length) for resistance data. 
Dashed lines denote 
the critical value of the 
DFA-indicator~\cite{livina07}. Red curves correspond 
to probability densities 
(kernel distributions of the times of projections of 
subsets of the indicators 
when critical value would be reached).
Panels correspond to the data in Fig.\ref{f2}.
}
\label{f4}
\end{figure}

\begin{figure}
\includegraphics[width=1.3\linewidth]{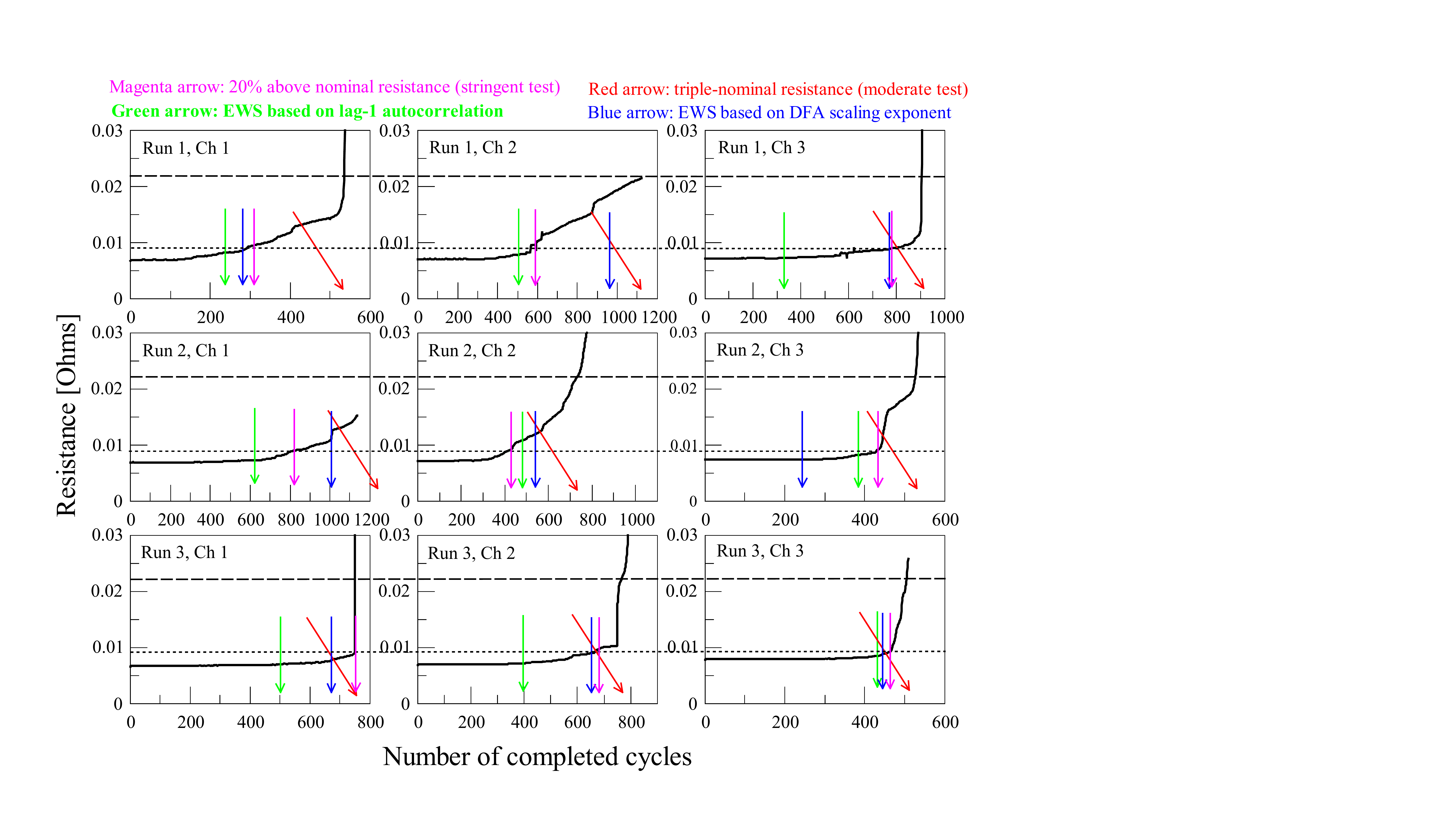}
\caption{Comparison of performance of two early 
warning signal indicators (green arrow for time stamp --- 
lag-1 ACF-indicator;
blue arrow --- DFA-indicator) and 
two conventional test thresholds (magenta arrow for time stamp ---
stringent test with 20\% resistance increase, denoted by dotted line;
red arrow ---
moderate test with triple-nominal resistance threshold,
denoted by dashed line).
Panels correspond to the data in Fig.\ref{f2}.
}
\label{f5}
\end{figure}

We first calculate the ACF-based indicator with 
uncertainty quantification based on varying window sizes
(between 1/4 and 3/4 of the data length)
and estimate the 
time of the early warning signals for them when 
the ACF-indicator reaches a high value of 0.9, as shown in 
Figure~\ref{f3}. 
In addition, we consider the average curve of the ACF-indicator 
and along this curve calculate linear extrapolation 
of the indicator to estimate when in future it would 
reach critical value~1 
(for DFA, the critical value is~1.5). 
By doing this, we obtain a set of possible times when the 
failure would happen, which forms a histogram --- 
this histogram is then used to generate the kernel density 
of the future failure times. The peak of such a kernel density 
is the most likely time of failure, statistically. 
We illustrate this in Figs.\ref{f3},\ref{f4} 
and also use this information in Fig.\ref{f5}.

We assume the real-time situation while moving with sliding windows  
along the time series and forming the indicators curves (this is what
happens when a time series is analysed real-time rather than in 
retrospect). 
To estimate the kernel distribution shown in the figure, 
we perform projections 
(linear extrapolations) of the indicators curves to 
obtain the statistics 
of the future state.

We also apply the DFA-indicator to assess early warning signals 
by  an alternative technique\ref{f4}. 
 In most cases, ACF shows earlier warnings than DFA.
This is caused by the difference between single-point ACF estimation
(lag-1) and the multiple-point DFA estimate (subset of the DFA curve 
in the time scale 10-100, as introduced in \cite{livina07}).
 
We then map the time points that can be seen in the rising 
indicators to the plot with the data, in which we also highlight 
where the electrical interconnect fails and goes open circuit, 
and observe that early warning signal indicators provide much 
earlier forewarning than the conventional technique (Figure~\ref{f5}).
Both early warning signal indicators provide 
earlier forewarning of the upcoming failure of units with 
critically rising resistance,  as compared with both stringent
(magenta arrow) and moderate (red arrow) threshold tests.
The stringent
test uses the criterion of 20\% increase of resistance~\cite{std},
whereas the moderate test uses triple-nominal resistance, which 
is obtained as the mean value of the initial nine estimates of 
resistance over first 50 cycles.
The locations of green and blue 
arrows are based on the peaks of kernel distributions in 
Figs.\ref{f3},\ref{f4}.

The variability of locations of early warning 
points in Fig.\ref{f5} is caused by different dynamics of the 
resistance time series: some of them fail more gradually, 
whereas others fail abruptly. Most likely, this is related 
to the material composition of the devices, which vary 
at mesoscopic level.

\phantom{.}
\vspace{30mm}
 
\section{Discussion}

We have applied early warning signal indicators to the power 
measurement data and, to our best knowledge, for the first time 
observed  an early forewarning of a failure in the 
 electric measurements. 
These techniques can be more accurate than bulk failure estimates 
because of  their application to a specific individual device, 
whereas compared with threshold-based detection, 
early warning signals have the advantage of indicating proximity
of failure in advance.

Machine learning techniques can infer the values 
of model parameters from the data. The proposed techniques 
estimate development of scaling properties of time series, which 
makes it similar to machine learning techniques --- however, 
this is done using windowed subsets of the same time series.
This makes it similar to bootstrapping and does not require 
large number of training datasets but rather utilises a single
time series. In real-time monitoring, the time series size 
increases with time, and the EWS indicators increase accordingly, 
until their values become critical. 

In the context of this study, the difference between 
transition (drift of a dynamical system) and bifurcation 
(change of the number of states) is not relevant. However, we 
note that in the observed resistance time series the critical 
behaviour is likely to be transitional. 

The methodology is generic and can be applied to other 
types of components. As those vary and have a wide range of
resistance levels, their conventional tests use specific thresholds.
For example, they combine per cent levels plus several ohms, 
or several events exceeding 1000  ohms for certain period, or
20\% resistance increase in five consequtive readings
\cite{std}. 
Our methodology is independent of such conditions, which 
is an advantage.  Moreover, it provides early warnings
signals of failures earlier than the conventional conditions with
both stringent (20\% excess of nominal value) and moderate 
(triple-nominal) thresholds (Fig.\ref{f5}).

 In terms of data processing, 
it is necessary to mention that 
any filtering affects autocorrelations, and therefore 
applying such data filter would distort the early warning 
signals in the data. 
For example, a low-pass filter would smooth the time series, thus 
increasing auto-correlations (dependencies of the 
close datapoints). This would mask the early warning 
signal and make the indicator oscillate at critical values 
without providing meaningful forewarning. High-pass filter
would decrease autocorrelations, thus reducing the values of the 
EWS indicators; in this case, however, it would be still possible 
to obtain early-warning signals from the trend in the indicators, 
but not from their values (i.e. they may not reach the critical values, 
but the trend will be present in the indicators). 

 The applied analysis does not depend on the types of the 
considered resistors as it is based on a 
generic stochastic model and is scalable to various 
levels of time series osciallations. Early warning signals 
are based on the changes in autocorrelations 
rather than on absolute values of  
fluctuations, by construction of indicators.

The devices whose resistance data we have analysed
are expected to be exploited 
in harsh conditions, such as in propulsion installations.
Unforeseen failure of such devices 
may cause life-threatening conditions, and therefore early 
warning signals may help avoid dangerous situations by means of  
timely replacement of aged components.

Although the techniques applied 
here provide advance early warnings that could be suitable for 
early safe replacement of endangered units, we understand that 
in industrial practice, economic considerations may dictate later 
predictive maintenance than is indicated by the proposed techniques. 
Finding the balance of early forewarning and further use of the 
unit undergoing the critical change may depend on the dynamics of 
the device and criticality of the unit in terms of safety.

\end{document}